\documentclass[prd,aps,preprintnumbers,showpacs,superscriptaddress,nofootinbib]{revtex4}

\usepackage{url}
\usepackage{amsmath}
\usepackage{array}
\usepackage{amssymb}
\usepackage{cancel}
\usepackage{epsfig}
\usepackage{graphicx}
\usepackage{psfrag}
\usepackage{slashed}

\newcommand{\be}{\begin{equation}}
\newcommand{\ee}{\end{equation}}
\newcommand{\bea}{\begin{eqnarray}}
\newcommand{\eea}{\end{eqnarray}}

\newcommand{\bm}[1]{\mbox{\boldmath $#1$}}

\newcommand{\st}{{\scriptscriptstyle T}}
\newcommand{\sL}{{\scriptscriptstyle L}}

\begin{document}

\preprint{NIKHEF 2015-007}
\title{Operator analysis of $p_T$-widths of TMDs}

\author{D. Boer}
\email{d.boer@rug.nl}
\affiliation{Van Swinderen Institute, University of Groningen, Nijenborgh 4,
NL-9747 AG Groningen, The Netherlands}

\author{M.G.A.~Buffing}
\email{m.g.a.buffing@vu.nl}
\affiliation{Nikhef and
Department of Physics and Astronomy, VU University Amsterdam,\\
De Boelelaan 1081, NL-1081 HV Amsterdam, the Netherlands}

\author{P.J.~Mulders}
\email{mulders@few.vu.nl}
\affiliation{Nikhef and
Department of Physics and Astronomy, VU University Amsterdam,\\
De Boelelaan 1081, NL-1081 HV Amsterdam, the Netherlands}

\begin{abstract}
Transverse momentum dependent (TMD) parton distribution functions (PDFs), TMDs for short, are defined as the Fourier transform of matrix elements of nonlocal combinations of quark and gluon fields. The nonlocality is bridged by gauge links, which for TMDs have characteristic paths (future or past pointing), giving rise to a process dependence that breaks universality. It is possible, however, to construct sets of universal TMDs of which in a given process particular combinations are needed with calculable, 
process-dependent, coefficients. This occurs for both T-odd and T-even TMDs, including also the {\it unpolarized} quark and gluon TMDs. This extends the by now well-known example of T-odd TMDs that appear with opposite sign in single-spin azimuthal asymmetries in semi-inclusive deep inelastic scattering or in the Drell-Yan process. In this paper we analyze the cases where TMDs enter multiplied by products of two transverse momenta, which includes besides the $p_T$-broadening observable, also instances with rank two structures. To experimentally demonstrate the process dependence of the latter cases requires measurements of second harmonic azimuthal asymmetries, while the $p_T$-broadening will require measurements of processes beyond semi-inclusive deep inelastic scattering or the Drell-Yan process. Furthermore, we propose specific quantities that will allow for theoretical studies of the process dependence of TMDs using lattice QCD calculations. 
\end{abstract}
\date{\today}

\maketitle

\section{Introduction}
When considering transverse momentum dependence of parton distribution functions (PDFs) one must account for process dependence, which is related to the color flow in the hard process and which is reflected in a process dependence of the gauge links (GLs) or Wilson lines that appear in the definition of the quark and gluon transverse momentum dependent (TMD) correlators~\cite{Collins:1981uw,Collins:1981uk,Brodsky:2002rv, Collins:2002kn,Belitsky:2002sm,Boer:2003cm,Bomhof:2006dp,Bomhof:2007xt,Buffing:2012sz,Buffing:2013kca}.
In a field theoretical framework in terms of quark and gluon field operators, these definitions involve nonlocal combinations of such fields and hence necessarily also gauge links. To `feel' the transverse momentum dependence they necessarily involve (covariant) derivatives or gluon field operators~\cite{Ellis:1982wd,Ellis:1982cd}. Probed in the right way, however, through specific azimuthal asymmetries, often necessarily in combination with transverse spin asymmetries, the effects of transverse momentum dependence are not suppressed by the high-energy scale in the process~\cite{Ralston:1979ys,Mulders:1995dh,Boer:1997nt}. This involves processes that are sensitive to an observable transverse momentum, which introduces another scale into the process, allowing the unsuppressed appearance of operators that would be suppressed with inverse powers of the hard scale in inclusive deep inelastic scattering. Such processes should be based on TMD factorization \cite{Collins:2011zzd,Echevarria:2012js} in order to link the observable to TMD correlators of quarks and gluons.

In the parametrization of TMD quark and gluon correlators one encounters the TMD distribution functions depending on $x$ and $p_\st^2$. In addition transverse momenta may appear explicitly, e.g.\ in the simplest case there are terms in the correlator with a linear proportionality to the (relative) transverse momentum of quarks with respect to parent or produced hadrons. Examples of new terms in this case include single-spin asymmetries that at the level of distribution functions are hard to obtain without transverse momentum effects. They can be traced to particular time-reversal-odd (T-odd) matrix elements involving gluonic pole matrix elements~\cite{Efremov:1981sh,Efremov:1984ip,Qiu:1991pp,Qiu:1991wg,Qiu:1998ia,Kanazawa:2000hz}. Starting with GL-dependent TMD correlators, such matrix elements appear with calculable process-dependent gluonic pole factors depending on the path of the GL~\cite{Qiu:1998ia,Bomhof:2004aw,Bacchetta:2005rm,Bomhof:2006ra,Bomhof:2007xt,Buffing:2012sz,Buffing:2013kca}. The path can be future or past-pointing dependent on the color flow in the hard process. This leads for instance to the prediction of the sign change (factors $\pm 1$) of comparable single-spin asymmetries in semi-inclusive deep inelastic scattering (SIDIS) or the Drell-Yan process (DY)~\cite{Brodsky:2002rv,Collins:2002kn,Belitsky:2002sm}.

In this paper, we focus on the operator structure of the matrix elements in situations where transverse momenta show up at the quadratic level, either as $p_\st^2$ or as a rank 2 tensor combination. The first one is relevant in studies involving the $p_\st^2$ dependence of the TMDs and its possible process dependence.
In the case of tensor combinations it leads to TMD functions in the parametrization that show up in the description of azimuthal asymmetries such as $\cos(2\varphi)$ or $\sin(2\varphi)$, where $\varphi$ is an appropriate azimuthal angle that can be defined in processes where at least two hadrons are involved, such as SIDIS or DY. 

The upshot of the paper is that transverse momentum dependence and the GL dependence in the description of TMDs in terms of matrix elements of nonlocal combinations of quark and gluon fields leads to a process dependence already for the rank zero TMDs, which are just the extension of the `standard' collinear PDFs with an extra argument. In that case it affects in particular the $p_\st$-width of the TMDs. Process dependence of unpolarized TMDs has already been considered at small $x$ in \cite{Dominguez:2011wm,Schafer:2013mza,Xiao:2010sp,Xiao:2010sa}. 
In this paper, we outline the underlying operator structure, i.e.\ the split into universal operator combinations and its process-dependent coefficients. Besides the $p_\st$-width of TMDs also situations are considered in which higher-rank effects in $p_\st$ show up, e.g.\ those involving the pretzelocity functions for transversely polarized quarks in a transversely polarized nucleon or in the description of linearly polarized gluons in an unpolarized nucleon. We link appropriate operators to a universal set of transverse momentum dependent distribution functions. 

In the analysis of operator structures it is convenient to use $p_\st$ moments of TMDs, which formally are ill-defined. The average partonic transverse momentum $\langle p_\st^2 \rangle$  is a theoretical quantity that beyond tree level requires a regularization and prescription in order to be well-defined. 
It is not directly observable. In this article we will not focus on this aspect and just view such transverse moments as idealized quantities, that are limits of well-defined Bessel moments for example~\cite{Boer:2011xd,Boer:2014bya}. Such Bessel moments are less sensitive to the large $p_\st$ behavior of the TMDs, but they are still scale dependent and under changes in scale they will involve operator mixing between quark and gluon operators, and even between operator combinations describing unpolarized and polarized gluons. We will come back to the latter at the end of section~\ref{sect:gluons}.

Assuming a proper definition in this way, and assuming TMD factorization, the $p_\st$ moments can be connected to experimental observables. 
Of course, in practice the isolation of the average partonic transverse momentum from experimental measurements is extremely challenging due to the multitude of contributing effects. For example, the transverse momentum imbalance of dijet, dimuon and diphoton pairs in hadron-hadron collisions has been used to extract the average partonic transverse momentum~\cite{Apanasevich:1997hm,Apanasevich:1998ki}. Dimuons and diphotons are primarily sensitive to quark contributions, but under scale changes also gluon contributions enter. Dijets probe the transverse momentum of both quarks and gluons already at the leading order. In the analysis the observed transverse momentum $Q_\st$ of the pair was divided into three parts: $\langle Q_\st^2 \rangle_{{\rm pair}}/2 = \langle p_\st^2 \rangle_{{\rm intrinsic}}+ \langle p_\st^2 \rangle_{{\rm soft}} + \langle p_\st^2 \rangle_{{\rm NLO}}$, based on the fact that the `intrinsic' transverse momentum distribution of partons beyond tree level is broadened by hard and soft emissions. Therefore, the theoretical description will require inclusion of ``$Y$-terms'' to bridge the low and high parts in the observable transverse momentum $Q_\st$. To reduce the sensitivity to the large $p_\st$ contributions, here one could also consider extracting the Bessel moments directly from experimental data. See Ref.\ \cite{Aghasyan:2014zma} for a first analysis of this kind. 

There are further complications in the experimental extractions. Although there is no contribution from fragmentation for these particular observables, the experimental transverse momentum resolution of the pair is a clear limiting factor and generally is of the same order or larger than the average partonic transverse momentum. In principle there are also contributions from spin correlation effects that were not accounted for in the experimental extraction \cite{Boer:2009nc,Pisano:2013cya}, and another effect not accounted for in the extractions, is the process dependent color flow factors arising from initial and/or final state interactions (ISI/FSI). In fact, due to the presence of both ISI and FSI in dijet production, TMD factorization of that process is not expected to be valid~\cite{Rogers:2010dm}. The extraction of the average partonic transverse momentum from the dijet imbalance in hadron-hadron collisions is therefore questionable, unless one can demonstrate that the spin correlation effects and the ISI/FSI effects are negligibly small. 

Despite the many complications in going from experimental measurements to trustworthy extractions of the $p_T$ widths of TMDs, it seems useful to investigate the possible effects of the color flow, and to find ways to assess the importance of such effects. The aim of this article is to shed light on this aspect and to propose ways to investigate it further quantitatively using lattice calculations, as we will discuss in section \ref{sect:lattice}.

\section{Definitions and parametrization of TMD correlators}
The quark and gluon TMD correlators in terms of matrix elements of quark fields~\cite{Collins:1981uw,Collins:1981uk} including the Wilson lines $U$ needed for color gauge invariance of the TMD case~\cite{Belitsky:2002sm,Boer:2003cm} are given by~\cite{Bomhof:2006dp,Bomhof:2007xt}
\bea
&&
\bm\Phi_{ij}^{[U]}(x,p_{\st};n)
=\int \frac{d\,\xi{\cdot}P\,d^{2}\xi_{\st}}{(2\pi)^{3}}
\,e^{ip\cdot \xi} \langle P{,}S\vert\overline{\psi}_{j}(0)
\,U_{[0,\xi]}\psi_{i}(\xi)\vert P{,}S\rangle\,\big|_{\xi\cdot n=0},
\label{e:operator}
\\&&
2x\,\bm\Gamma^{[U,U^\prime]\,\mu\nu}(x,p_\st;n) ={\int}\frac{d\,\xi{\cdot}P\,d^2\xi_\st}{(2\pi)^3}\ e^{ip\cdot\xi}
\,\langle P{,}S\vert\,F^{n\mu}(0)\,U_{[0,\xi]}^{\phantom{\prime}}\,F^{n\nu}(\xi)\,U_{[\xi,0]}^\prime\,\vert P{,}S\rangle\big|_{\xi\cdot n=0}
\eea
(color summation or tracing implicit), where we use the Sudakov decomposition $p^{\mu}=x P^{\mu}+p_{\st}^{\mu}+\sigma n^{\mu}$ for the momentum $p^{\mu}$ of the produced quark or gluon. In the TMD case, there are for a spin 1/2 nucleon in general eight leading contributing terms in the parametrization of the TMD correlator~\cite{Mulders:1995dh,Bacchetta:2006tn}, but as already explained in the Introduction we focus only on specific examples, namely unpolarized quarks in an unpolarized nucleon or transversely polarized quarks in a transversely polarized nucleon,
\begin{eqnarray}
\bm\Phi^{[U]}(x,p_{\st};n)&=&\bigg\{
f^{[U]}_{1}(x,p_\st^2)
+h^{[U]}_{1}(x,p_\st^2)\,\gamma_5\,\slashed{S}_{\st}
+h_{1T}^{\perp [U]}(x,p_\st^2)\,\frac{p_{\st\alpha\beta}S_\st^{\{\alpha}\gamma_\st^{\beta\}}\gamma_5}{2M^2}
\bigg\}\frac{\slashed{P}}{2} .
\label{e:par-quark2}
\end{eqnarray}
The spin vector is parametrized as 
$S^\mu = S_{\sL}P^\mu + S^\mu_{\st} + M^2\,S_{\sL}n^\mu$ (with in our case $S_\sL$ not needed). The symmetric traceless second rank tensor is $p_\st^{\alpha\beta} = p_\st^\alpha\,p_\st^{\beta} - \frac{1}{2}p_\st^2\,g_\st^{\alpha\beta}$. In this parametrization this traceless tensor is used. Often one also encounters the expression in which just the symmetric combination $p_\st^\alpha p_\st^\beta$ is used. In that case one has a trace term that in the above expression is absorbed into $h_1$ introducing the combination $h_1^{[U]} = h_{1T}^{[U]} + h_{1T}^{\perp [U](1)}$. The notation $h_{1T}^{\perp (1)}$ indicates weighting with powers of $-p_\st^2/2M^2 = \bm p_\st^2/2M^2$, of which the general definition reads:
\be
f_{\ldots}^{(n)}(x,p_\st^2) = \left(\frac{-p_\st^2}{2M^2}\right)^n\,f_{\ldots}(x,p_\st^2),
\label{e:transversemoments}
\ee
and which upon integration are referred to as transverse moments. 
For gluons one has for the leading correlator in a polarized nucleon in general eight GL-dependent functions. We limit ourselves to unpolarized nucleons with for gluons the parametrization~\cite{Mulders:2000sh}
\bea
&&
2x\,\bm\Gamma^{\mu\nu}(x,p_\st) =
-g_T^{\mu\nu}\,f_1^{g [U,U^\prime]}(x,p_\st^2)
+\frac{p_T^{\mu\nu}}{M^2}\,h_1^{\perp g[U,U^\prime]}(x,p_\st^2),
\label{e:GluonCorr-par}
\eea
which for gluons is sufficient to illustrate the complications when a product of two transverse momenta is involved. The (formal) integrated correlators are given by
\bea
&&
\bm\Phi_{ij}(x)
=\int \frac{d\,\xi{\cdot}P}{2\pi}e^{ip\cdot \xi} 
\langle P{,}S\vert\overline{\psi}_{j}(0)U_{[0,\xi]}^{[n]}
\psi_{i}(\xi)\vert P{,}S\rangle\,\big|_{\xi\cdot n=0,\xi_{\st}=0},
\label{e:col-operator-quark}
\\&&
2x\,\bm\Gamma^{\mu\nu}(x) ={\int}\frac{d\,\xi{\cdot}P}{(2\pi)}\ e^{ip\cdot\xi}
\,\langle P{,}S\vert\,F^{n\mu}(0)\,U_{[0,\xi]}^{[n]}\,F^{n\nu}(\xi)\,U_{[\xi,0]}^{[n]}\,\vert P{,}S\rangle\big|_{\xi\cdot n=0,\xi_{\st}=0} ,
\label{e:col-operator-gluon}
\eea
where the GLs are reduced to a unique straight-line GL which runs from $0$ to $\xi$ along $n$, thus removing the link dependence. The relevant parametrizations for the collinear quark and gluon correlators in the examples (Eqs.~\ref{e:par-quark2} and \ref{e:GluonCorr-par}) which we study in this paper are
\bea
&&
\bm\Phi(x) = \bigg\{f_1(x) +h_1(x)\,\gamma_5\,\slashed{S}_{\st}\bigg\}\frac{\slashed{P}}{2},
\label{e:collinear-quark}
\\&&
2x\,\bm\Gamma^{\mu\nu}(x) = -g_\st^{\mu\nu}\,f_1^g(x).
\label{e:collinear-gluon}
\eea
Here we do not discuss any complicating issues associated with scale dependence of the distribution functions and with convergence of the $p_\st$ integrals. These are addressed briefly in \ref{sect:lattice} and more extensively in \cite{Collins:2003fm,Boer:2014bya}. In the next sections, we make the GL dependence in the TMD functions like $f_1^{[U]}(x,p_\st^2)$ for quarks or $f_1^{g[U,U^\prime]}(x,p_\st^2)$ for gluons explicit.

\section{Quark correlators of rank two in a nucleon target}
In this section, we study the quark correlator in Eq.~\ref{e:par-quark2} and we will identify the $p_\st$ dependence with specific operators. For this we notice that the $p_\st$-integrated collinear correlator,
\bea
\int d^2p_\st\ \bm\Phi^{[U]}(x,p_\st) & = & \bm\Phi(x), 
\label{e:quarkcollinear}
\eea
is GL-independent. Likewise one can consider higher transverse moments. Using the starting operator expression in Eq.~\ref{e:operator} one finds rank two collinear correlators~\cite{Boer:2003cm,Bacchetta:2005rm},
\bea
&&
\int d^2p_\st\ \frac{p_\st^\alpha p_\st^\beta}{M^2}\,\bm\Phi^{[U]}(x,p_\st) 
= \bm\Phi_{\partial\partial}^{\alpha\beta}(x) 
+ \sum_{c=1}^{2}C_{GG,c}^{[U]}\,\bm\Phi_{GG,c}^{\alpha\beta}(x),
\label{e:quarktransversemoments}
\eea
which are symmetric but not traceless. Subtracting its trace we get
\bea
&&
\int d^2p_\st\ \frac{p_\st^{\alpha\beta}}{M^2}\,\bm\Phi^{[U]}(x,p_\st) 
= \bm\Phi_{\partial\partial}^{\alpha\beta}(x)
-\frac{1}{2}\,g_\st^{\alpha\beta}\,\bm\Phi_{\partial{\cdot}\partial}(x) 
+ \sum_{c=1}^{2}C_{GG,c}^{[U]}\left(\bm\Phi_{GG,c}^{\alpha\beta}(x)
-\frac{1}{2}\,g_\st^{\alpha\beta}\,\bm\Phi_{G{\cdot}G,c}(x)\right).
\label{e:quarktransversemoments-rank}
\eea
These are relations involving on the right-hand side correlators with specific operator matrix elements, among them two ($c = 1,2$) gluonic pole matrix elements $\bm\Phi_{GG,\,c}$, multiplied with gluonic pole factors $C_{GG,c}^{[U]}$. The explicit operator structure of these matrix elements before doing the $p_\st$-integration is given by
\bea
&&
\bm\Phi_{\hat O,ij}^{[U]}(x,p_{\st})
=\int \frac{d\,\xi{\cdot}P\,d^{2}\xi_{\st}}{(2\pi)^{3}}
\,e^{ip\cdot \xi} \langle P{,}S\vert\overline{\psi}_{j}(0)
\,U_{[0,\xi]}\hat O(\xi)\psi_{i}(\xi)\vert P{,}S\rangle\,\Big|_{\xi\cdot n=0},
\label{e:twistoperatorquark}
\eea
where the $\hat O(\xi)$ operators are rank two combinations of $i\partial_\st(\xi) = iD_\st^\alpha(\xi) - A_\st^\alpha(\xi)$ and $G^\alpha(\xi)$, defined in a color gauge invariant way (thus including GLs),
\bea
&&A_{\st}^{\alpha}(\xi)=\frac{1}{2}\int_{-\infty}^{\infty}
d\eta{\cdot}P\ \epsilon(\xi{\cdot}P-\eta{\cdot}P)
\,U_{[\xi,\eta]}^{[n]} F^{n\alpha}(\eta)U_{[\eta,\xi]}^{[n]}, 
\label{e:defA} 
\\
&&G^{\alpha}(\xi)=\frac{1}{2}\int_{-\infty}^{\infty}
d\eta{\cdot}P\ U_{[\xi,\eta]}^{[n]}F^{n\alpha}(\eta)
U_{[\eta,\xi]}^{[n]},
\label{e:defG}
\eea
with $\epsilon (\zeta)$ being the sign function. Note that $G^{\alpha}(\xi)$ = $G^{\alpha}(\xi_\st)$ does not depend on $\xi{\cdot}P$, implying in momentum space $p\cdot n = p^+ = 0$, hence the name gluonic pole matrix elements. To be precise the correlator $\bm\Phi_{GG,c}$ contains the operator $G^\alpha G^\beta/M^2$. Including the quark fields in the correlator, the different color possibilities correspond to the color singlet combinations ${\rm Tr}(\psi\overline\psi GG)$ ($c=1$) or ${\rm Tr}(\psi\overline\psi){\rm Tr}(GG)$ ($c=2$). The other correlator in Eq.~\ref{e:quarktransversemoments}, $\bm\Phi_{\partial\partial}$, contains the operator $(i\partial_\st^\alpha)(i\partial_\st^\beta)/M^2$. The collinear correlators $\bm\Phi_{\partial\partial}$ and $\bm\Phi_{GG,c}$ are universal. 
In our particular examples, both of them are rank 2 symmetric correlators, but in general these are not traceless.

In the next step, we are going to study the parametrization of the correlators. For this we will distinguish among the terms in the parametrization with different numbers of transverse momenta, i.e.\ terms with definite rank. To illustrate the procedure, it is sufficient to distinguish in our example the symmetric terms up to rank two,
\bea
&&\bm\Phi^{[U]}(x,p_\st) =
\widetilde\Phi^{[U]}(x,p_\st^2) + \widetilde\Phi^{\alpha[U](1)}_{\,\alpha}(x,p_\st^2) 
+ \frac{p_{\st \alpha\beta}}{M^2}
\,\widetilde\Phi^{\alpha\beta [U]}(x,p_\st^2),
\label{e:basicexpansion-quark}
\eea
parametrized as
\bea
&&
\widetilde\Phi^{[U]}(x,p_{\st}^2) + \widetilde\Phi^{\alpha[U](1)}_{\,\alpha}(x,p_\st^2) 
= \bigg\{ f^{[U]}_{1}(x,p_\st^2)
+ h^{[U]}_{1}(x,p_\st^2)\,\gamma_5\,\slashed{S}_{\st}
\bigg\}\frac{\slashed{P}}{2},
\label{e:rank0}
\\&&
\widetilde\Phi^{\alpha\beta [U]}(x,p_{\st}^2)
+ \frac{1}{2}\,g_\st^{\alpha\beta}\,\widetilde\Phi^{\gamma[U]}_{\,\gamma}(x,p_\st^2)
= h_{1T}^{\perp [U]}(x,p_\st^2)\,\frac{S_{\st}^{\{\alpha}\gamma_\st^{\beta\}}\gamma_5 
+ g_\st^{\alpha\beta}\gamma_5 \slashed{S}_\st}{2}
\,\frac{\slashed{P}}{2},
\label{e:rank2}
\eea
where the tensor $\widetilde\Phi^{\alpha\beta}$ is symmetric but not traceless. Note that the weighted trace term $\widetilde\Phi_{\,\alpha}^{\alpha\, [U](1)}(x,p_{\st}^2)$ then contributes to the parametrization of the rank 0 part, while the trace term also has to be subtracted in the rank 2 part. This suggests absorption of the first two terms in Eq.~\ref{e:rank0} into one term, $\widetilde\Phi^{[U]} + \widetilde\Phi^{\alpha[U](1)}_{\,\alpha}\longrightarrow \widetilde\Phi^{[U]}$, which indeed is possible if we realize that the terms are GL-dependent anyway. We note that the notation using tildes above the quark and gluon correlators in this expansion sometimes slightly differs from that in previous studies. In this paper, we simply use tildes for particular definite rank parts in the parametrization, which thus only depend on $x$ and $p_\st^2$. 

The $p_\st$-integrated and weighted collinear correlators in Eqs.~\ref{e:quarkcollinear}, \ref{e:quarktransversemoments} and \ref{e:quarktransversemoments-rank} are independent of the GL. The GL dependence is contained in the gluonic pole factors. It is then natural to also split the rank 2 contributions in Eq.~\ref{e:basicexpansion-quark} and start with the expansion
\bea
\bm\Phi^{[U]}(x,p_\st) & = &
\widetilde\Phi(x,p_\st^2) 
+\frac{p_{\st \alpha\beta}}{M^2}
\,\widetilde\Phi_{\partial\partial}^{\alpha\beta}(x,p_\st^2)
\nonumber \\ && \mbox{} 
+\sum_{c=1}^{2}C_{GG,c}^{[U]}\left[\widetilde\Phi_{G{\cdot}G,c}^{(1)}(x,p_\st^2) + \frac{p_{\st \alpha\beta}}{M^2}
\,\widetilde\Phi_{GG,c}^{\alpha\beta}(x,p_\st^2)\right] + \ldots\ .
\label{e:expansion}
\eea
Also in this expansion we thus must account for the trace terms appearing as (universal) matrix elements $\widetilde\Phi_{\partial{\cdot}\partial}(x,p_\st^2)$ and $\widetilde\Phi_{G{\cdot}G}(x,p_\st^2)$ in the rank 0 part. In the remaining part indicated by the dots only terms with additional $G{\cdot}G$ will appear since, for the $p_\st$ dependence, rank-2 structures are the highest ones for quark matrix elements in a spin 1/2 hadron. Because $\Phi(x)$ and $\widetilde\Phi_{\partial{\cdot}\partial}(x)$ are GL-independent we can make the substitution $\widetilde\Phi(x,p_\st^2) + \widetilde\Phi^{\ (1)}_{\partial{\cdot}\partial}(x,p_\st^2)\longrightarrow \widetilde\Phi(x,p_\st^2)$. The rank 0 trace part of the gluonic pole correlator, however, must be kept and parametrized in analogy to the rank 0 part $\Phi(x,p_\st^2)$. It cannot contribute to the integrated correlators, thus requiring that upon integration $\widetilde\Phi_{G{\cdot}G,c}^{(1)}(x) = 0$. Following the naming convention in Ref.~\cite{Buffing:2012sz} for the universal functions getting a superscript $(A)$ for functions in $\widetilde\Phi_{\partial\partial}$ and a superscript $(Bc)$ for functions in $\widetilde\Phi_{GG,c}$, this leads to the following parametrization in terms of universal functions,
\bea
&&
\widetilde\Phi(x,p_\st^2)
=\bigg\{ f_{1}(x,p_\st^2) +h_1(x,p_\st^2)\,\gamma_5\,\slashed{S}_{\st}
\bigg\}\frac{\slashed{P}}{2},
\\&&
\widetilde\Phi_{\partial\partial}^{\alpha\beta}(x,p_\st^2) 
+ \frac{1}{2}\,g_{\st}^{\alpha\beta}\,\widetilde\Phi_{\partial{\cdot}\partial}(x,p_\st^2)
= h_{1T}^{\perp (A)}(x,p_\st^2)
\,\frac{S_{\st}^{\{\alpha}\gamma_\st^{\beta\}}\gamma_5 
+ g_\st^{\alpha\beta}\gamma_5 \slashed{S}_\st}{2}
\,\frac{\slashed{P}}{2},
\\&&
\widetilde\Phi_{G{\cdot}G,c}^{(1)}(x,p_\st^2)
= \bigg\{\delta f_1^{(Bc)}(x,p_\st^2) 
+ \delta h_{1}^{(Bc)}(x,p_\st^2)\,\gamma_5\,\slashed{S}_{\st}
\bigg\}\frac{\slashed{P}}{2},
\nonumber \\&&
\widetilde\Phi_{GG,c}^{\alpha\beta}(x,p_\st^2)
+\frac{1}{2}\,g_\st^{\alpha\beta}\,\widetilde\Phi_{G{\cdot}G,c}(x,p_\st^2)
= h_{1T}^{\perp (Bc)}(x,p_\st^2)
\,\frac{S_{\st}^{\{\alpha}\gamma_\st^{\beta\}}\gamma_5 
+ g_\st^{\alpha\beta}\gamma_5 \slashed{S}_\st}{2}
\,\frac{\slashed{P}}{2},
\label{e:highrank}
\eea
with $\delta f_1^{(Bc)}(x) = \delta h_1^{(Bc)}(x) = 0$. Making the trace term $\Phi_{\partial{\cdot}\partial}$ explicit by introducing additional functions $\delta f_1^{(A)}$ and $\delta h_1^{(A)}$ would be overcomplete\footnote{Note the missing factor of $1/2$ for the terms in the last line of Eq.~29 in Ref.~\cite{Buffing:2012sz}. Also the order of some Dirac matrices in the Eqs. 31-33 of the same reference should be reversed.}.

Returning to our starting GL-dependent parametrization in Eq.~\ref{e:par-quark2}, we have now constructed the universal set of functions, including $f_1$ and $h_{1T}$, their $p_\st$-width effects $\delta f_1^{(Bc)}$ and $\delta h_{1}^{(Bc)}$ and three pretzelocity functions that give the GL-dependent distribution functions,
\bea
f_1^{[U]} &=& f_1 + \sum_{c=1}^{2}C_{GG,c}^{[U]}\,\delta f_1^{(Bc)} + \ldots\ ,
\\
h_1^{[U]} &=& h_1 + \sum_{c=1}^{2}C_{GG,c}^{[U]}\,\delta h_1^{(Bc)} + \ldots\ ,
\\
h_{1T}^{\perp [U]} &=& h_{1T}^{\perp (A)} 
+ \sum_{c=1}^{2}C_{GG,c}^{[U]}\, h_{1T}^{\perp (Bc)} + \ldots\ .
\eea
Note that all functions in these equation depend on $x$ and $p_\st^2$. For the pretzelocity distributions, this has been discussed in \cite{Buffing:2012sz}, but the broadening effects parametrized with two functions $\delta f_1^{(Bc)}$ is new. In principle traces of higher-rank tensors will give broadening effects coming with color factors $C_{GGGG,c}^{[U]}$, etc. These will not lead to new functions, but just to additional process-dependent broadening effects $\delta\delta f_1$, $\delta\delta h_1$ and also broadening effects $\delta h_{1T}^{\perp (1)}$ in the pretzelocity functions. These satisfy $\delta\delta f_1(x) = \delta\delta f_1^{(1)} = 0$, $\delta\delta h_1(x) = \delta\delta h_1^{(1)} = 0$ and $\delta h_{1T}^{\perp (1)}(x) = 0$, etc. Such functions, however, only start playing a role in processes with rather complex color flow. Furthermore they just represent modulations in the $p_\st^2$ dependence. Comparing $\delta\delta f_1$
with $\delta f_1$ and $f_1$, respectively, such modulations are expected to correspond to short distance
effects in impact parameter space in the matrix elements. We expect these to become smaller and hope that this can be studied in lattice studies as discussed in Section~\ref{sect:lattice}.

\section{Gluon correlators of rank 2 in a nucleon target\label{sect:gluons}}
As the second example, we consider the gluon correlator in an unpolarized nucleon with the GL-dependent parametrization given in Eq.~\ref{e:GluonCorr-par}. As for quarks the integrated and weighted results for gluon correlators give two types of matrix elements, 
\bea
&&
\int d^2p_\st\ \bm\Gamma_{\quad}^{\mu\nu [U,U^\prime]}(x,p_\st) = \bm\Gamma_{\quad}^{\mu\nu}(x),
\\&&
\int d^2p_\st\ \frac{p_\st^\alpha p_\st^\beta}{M^2}\,\bm\Gamma_{\quad}^{\mu\nu [U,U^\prime]}(x,p_\st) 
= \bm\Gamma_{\quad\,\partial\partial}^{\mu\nu;\alpha\beta}(x) 
+ \sum_{c=1}^{4}C_{GG,c}^{[U,U^\prime]}\,\bm\Gamma_{\quad\,GG,c}^{\mu\nu;\,\alpha\beta}(x).
\eea
For the gluonic pole correlators the index $c$ now runs over 4 possibilities involving different color tracings. Including the two gluon fields in the correlator $\bm\Gamma$ (denoted with $F$) the relevant color structures are ${\rm Tr}(F[G,[G,F]])$, ${\rm Tr}(F\{G,\{G,F\}\})$, ${\rm Tr}(FF){\rm Tr}(GG)$ and ${\rm Tr}(FG){\rm Tr}(FG)$, respectively. These transverse moments can be used to identify universal functions after introducing a parametrization in terms of definite rank,
\bea
&&\bm\Gamma_{\quad}^{\mu\nu[U,U^\prime]}(x,p_\st) =
\widetilde\Gamma_{\quad}^{\mu\nu}(x,p_\st^2) 
+ \frac{p_{\st \alpha\beta}}{M^2}
\,\widetilde\Gamma_{\quad\,\partial\partial}^{\mu\nu;\,\alpha\beta}(x,p_\st^2)
\nonumber \\&&
\mbox{}\hspace{3.3cm} + \sum_{c=1}^4 C_{GG,c}^{[U,U^\prime]}\left[
\widetilde\Gamma_{\quad\,G{\cdot}G,c}^{\mu\nu\ (1)}(x,p_\st^2) 
+ \frac{p_{\st \alpha\beta}}{M^2}
\,\widetilde\Gamma_{\quad\,GG,c}^{\mu\nu;\,\alpha\beta}(x,p_\st^2)\right] + \ldots\ ,
\eea
with $\widetilde\Gamma_{\quad\partial{\cdot}\partial}^{\mu\nu\ (1)}(x,p_\st^2)$ absorbed in $\widetilde\Gamma^{\mu\nu}(x,p_\st^2)$, just as we did for the quark correlator in Eq.~\ref{e:expansion}. In the parametrization of the gluonic pole term a trace term $\widetilde\Gamma_{\quad\,G{\cdot}G,c}^{\mu\nu\ (1)}(x,p_\st^2)$ is needed which must satisfy $\widetilde\Gamma_{\quad\,G{\cdot}G,c}^{\mu\nu\ (1)}(x) = 0$. With this expansion we can express the correlators in a universal set of TMDs depending on $x$ and $p_\st^2$,
\bea
&&
\widetilde\Gamma_{\quad}^{\mu\nu}(x,p_\st^2) 
= -g_\st^{\mu\nu}\,f^{g}_{1}(x,p_\st^2),
\\ &&
\widetilde\Gamma_{\quad\,\partial\partial}^{\mu\nu;\,\alpha\beta}(x,p_\st^2)
+ \frac{1}{2}\,g_\st^{\alpha\beta}\,\widetilde\Gamma_{\quad\,\partial{\cdot}\partial}^{\mu\nu}(x,p_\st^2)
= \frac{g_\st^{\mu\{\alpha}g_\st^{\beta\}\nu}-g_\st^{\alpha\beta}\,g_\st^{\mu\nu}}{2}\,h_{1}^{\perp g(A)}(x,p_\st^2),
\\&&
\widetilde\Gamma_{\quad\,G{\cdot}G,c}^{\mu\nu;\,(1)}(x,p_\st^2)
= -g_\st^{\mu\nu}\,\delta f_{1}^{g (Bc)}(x,p_\st^2),
\\&&
\widetilde\Gamma_{\quad\,GG,c}^{\mu\nu;\,\alpha\beta}(x,p_\st^2)
+\frac{1}{2}\,g_\st^{\alpha\beta}\,\widetilde\Gamma_{\quad\,G{\cdot}G,c}^{\mu\nu}(x,p_\st^2)
=\frac{g_\st^{\mu\{\alpha}g_\st^{\beta\}\nu}-g_\st^{\alpha\beta}\,g_\st^{\mu\nu}}{2}\,h_{1}^{\perp g(Bc)}(x,p_\st^2),
\eea
where the $p_\st^2$-integrated function $\delta f_1^{g}(x)$ = 0.
Returning to the GL-dependent functions in Eq.~\ref{e:GluonCorr-par} we find
\bea
&&
f_1^{g [U,U^\prime]} = f_1^{g} 
+ \sum_{c=1}^{4} C_{GG,c}^{[U,U^\prime]}\,\delta f_1^{g (Bc)} + \ldots\ ,
\\&&
h_1^{\perp g [U,U^\prime]} = h_1^{\perp g (A)} 
+ \sum_{c=1}^{4} C_{GG,c}^{[U,U^\prime]}\, h_1^{\perp g (Bc)} + \ldots\ .
\eea

We already commented in the Introduction on the fact that the UV behavior of the functions requires care including operator mixture. The function $h_1^{\perp g}$ is also a good example of the situation in which the large $p_\st$ behavior of the function involves convolutions with functions corresponding to a different operator structure, that is in this context operators of a different rank. In this case it is the unpolarized gluon distribution that in the evolution equation also contributes to the large $p_\st$ behavior of $h_1^{\perp g}$ as already pointed out in Ref.~\cite{Catani:2010pd,Catani:2011kr}.

\section{Gauge link dependence in lattice studies\label{sect:lattice}}
TMDs are GL-dependent, but as the gauge links can be calculated for a particular hard process, one can express the GL dependence in gluonic pole factors. This will affect all TMDs, even the unpolarized quark and gluon TMDs causing a process dependence in their $p_\st^2$ behavior and leading to nonuniversal $T$-even functions. To probe this process dependence in case of the unpolarized quark TMDs, one has to go beyond the comparison between relatively simple processes like SIDIS and DY, since the gluonic pole factors are both unity for simple future- and past-pointing gauge links,
\bea
&&
f_1^{[+]}(x,p_\st^2) = f_1^{[-]}(x,p_\st^2) = f_1(x,p_\st^2) + \delta f_1^{(B1)}(x,p_\st^2).
\eea
In hadron-hadron scattering to hadronic final states one will find effects of unpolarized quarks coming from (among others) the correlators $\Phi^{[\Box +]}$ and $\Phi^{[(\Box)+]}$. These gauge links are more
complex than the simple staples $U^{[\pm]}_{[0,\xi]}$ which are consecutive Wilson lines
$U^{[\pm]}_{[0,\xi]} = U^{[n]}_{[0^-,\pm\infty^-]}U^{T}_{[0_\st,\xi_\st]}U^{[n]}_{[\pm\infty^-,\xi^-]}$ along
minus direction or in the transverse direction at plus or minus light-like infinity. The GL $U^{[\Box +]}$ indicates a 
future-pointing GL that loops around once more, 
$U^{[\Box +]}_{[0,\xi]} = U^{[+]}_{[0,\xi]}U^{[-]}_{[\xi,0]}U^{[+]}_{[0,\xi]}$, 
while $U^{[(\Box)+]}$ indicates a future-pointing GL $U^{[+]}$ and an additional traced GL 
$U^{[(\Box)]} 
= {\rm Tr}(U^{[+]}_{[0,\xi]}U^{[-]}_{[\xi,0]})/N_c$~\cite{Buffing:2011mj,Buffing:2012sz,Buffing:2013kca}. 
We find in these cases\footnote{Compared to Ref.~\cite{Buffing:2012sz} there is a redefinition of gluonic pole matrix elements and gluonic pole coefficients, leaving the product of the two unchanged.}
\bea
&&
f_1^{[\Box +]}(x,p_\st^2) = f_1(x,p_\st^2) + 9\,\delta f_1^{(B1)}(x,p_\st^2),
\\&&
f_1^{[(\Box) +]}(x,p_\st^2) = f_1(x,p_\st^2) + \delta f_1^{(B1)}(x,p_\st^2) + \delta f_1^{(B2)}(x,p_\st^2).
\eea
We emphasize once more that the collinear integrated functions are the same in all cases. It should be mentioned though that strictly speaking one should not consider the collinear parton distribution functions as integrals over TMDs, as the latter require regularization of rapidity or light-cone divergences, and thereby do not necessarily lead to unique answers for integrals \cite{Collins:2003fm}. In general, higher transverse moments, including the average transverse momentum $\langle p_\st^2 \rangle \equiv 2 M^2 f_1^{(1)}(x)$, will diverge too, simply because the perturbative power law tail of the TMDs will not fall off sufficiently fast. For the purpose of regularizing this kind of divergence, a generalization of the weighting with powers of transverse momentum was suggested in~\cite{Boer:2011xd}, the so-called Bessel weighting. Bessel moments can be given as derivatives of Fourier transformed TMDs $\tilde f(x, \boldsymbol{b}_T^2)$ in impact parameter space,
\be
\tilde f^{(n)}(x, \boldsymbol{b}_T^2) = n!\left( -\frac{2}{M^2}\partial_{ \boldsymbol{b}_T^2} \right)^n \ \tilde f(x, \boldsymbol{b}_T^2). 
\ee
In the limit $b_T \to 0$ the conventional transverse moments are retrieved, including their divergences. The Bessel weighting regularized version of the average transverse momentum is given by~\cite{Boer:2014bya}:
\be
\langle p_\st^2 \rangle(x, \boldsymbol{b}_T^2) \equiv 2 M^2 \tilde f_1^{(1)}(x, \boldsymbol{b}_T^2) = 4\pi \int d |\boldsymbol{p}_T| \frac{|\boldsymbol{p}_T|^2}{|\boldsymbol{b}_T|} J_1(|\boldsymbol{b}_T||\boldsymbol{p}_T|)\ f_1(x, {\boldsymbol{p}_T^2} ).
\ee
Upon taking Mellin moments, this quantity can be evaluated on the lattice~\cite{Musch:2011er,Engelhardt:2014wra}, which suggests a lattice study of the GL dependence of $\tilde{f}_1^{(1) {[U]} }(x, \boldsymbol{b}_T^2)$.
For this purpose one can for example consider regularized versions of ratios such as
\bea
&& \frac{f_1^{(1) [\Box +]}(x)}{f_1^{(1) [+]}(x)} = 1 + 8 \frac{R_1}{1+ R_1},
\quad\quad\quad\quad \frac{f_1^{(1) [(\Box)+]}(x)}{f_1^{(1) [+]}(x)} = 1 + \frac{R_2}{1+ R_1}, 
\eea 
where $R_c \equiv \delta f_1^{(1)(Bc)}(x)/ f_1^{(1)}(x)$. The Bessel-weighted generalizations of these ratios can be evaluated on the lattice, e.g.
\be
\frac{\tilde f_1^{[1] (1) [\Box +]}(\boldsymbol{b}_T^2;\mu,\zeta)}{\tilde f_1^{[1] (1) [+]}(\boldsymbol{b}_T^2;\mu,\zeta)} = \frac{\langle P\vert\overline{\psi}(0, 0_\st)\gamma^+\,U^{[+]}_{[0,b]}U^{[-]}_{[b,0]} U^{[+]}_{[0,b]}\,
\psi(0,b_T)\vert P \rangle}{\langle P\vert\overline{\psi}(0,0_\st)\gamma^+
\,U^{[+]}_{[0,b]}\, \psi(0,b_T)\vert P \rangle}, 
\ee
where the first superscript $[1]$ refers to the lowest Mellin moment $n=1$ and $(1)$ to the Bessel-moment given above. These ratios offer a way to quantify the importance of gauge loops and of the process dependence of $f_1$. Although it is not expected that the ratios are scale-independent, some of the scale dependence (of both $\mu$ and $\zeta$) may cancel in the ratio~\cite{Boer:2011xd}. But even information about them at some fixed scales would already be very interesting. Any deviation from unity indicates first of all the relevance of gluonic pole matrix elements and second, the effect of the flux through a Wilson loop. How the latter changes with $b$ is especially interesting. If the effect of an additional winding in the Wilson line enclosing the entire flux through the plane does not affect the average transverse momentum squared significantly, then one can conclude that the process dependence is not important to take into account for the unpolarized T-even distributions. Extensions to other T-even and T-odd functions are straightforward.

\section{Discussion and conclusions}
Even if quark TMDs in hadron-hadron scattering processes involve a combination of correlators $\Phi^{[\Box +]}$, $\Phi^{[(\Box)+]}$, and some others with more complex gauge link structures, the consequences for process dependence of the unpolarized quark TMD $f_1^{[U]}$ will most likely be hard to investigate experimentally. This will be looked at more carefully in a future study, which also includes other TMDs such as the T-even and T-odd rank 1 functions. Lattice calculations, however, could offer possibilities to investigate the gauge link dependence.
For the gluon case, the situation may experimentally be somewhat less complicated, while the lattice calculation will be much more demanding in that case. The unpolarized gluon TMD $f_1^{g [U,U^\prime]} $ depends on two links and again the two simplest processes are sensitive to the same function: production of a colorless final state in hadron-hadron scattering, such as in the case of the Drell-Yan process or Higgs production, probes $f_1^{g[-,-]}$~\cite{Boer:2013fca,Echevarria:2015uaa}, while $c\overline c$ production in SIDIS probes $f_1^{g[+,+]}$~\cite{Pisano:2013cya}. Like $f_1^{[+]}= f_1^{[-]}$, it turns out that $f_1^{g[-,-]}=f_1^{g[+,+]}$, which follows from a $P$ and $T$ transformation. For T-odd TMDs this equality does not hold~\cite{Buffing:2011mj}, e.g.\ for the gluon Sivers function $f_{1T}^{\perp g[-,-]} \neq f_{1T}^{\perp g [+,+]}$. Processes in which other $f_1^{g [U,U^\prime]}$ functions contribute are for instance $p p \to H \, {\rm jet}\, X$ and $p p \to \gamma\, {\rm jet}\, X$. In $p p \to H \, {\rm jet}\, X$~\cite{Boer:2014lka} the subprocesses $q\bar{q} \to H g$, $qg \to H q$, and $gg \to Hg$ contribute, but when the momentum fractions of both initial partons is sufficiently small, the latter dominates, offering the possibility of accessing a single $f_1^{g [U,U^\prime]}$, which can be evaluated following the procedures as outlined in Ref.~\cite{Bomhof:2006dp}. A simpler way to probe a different $f_1^{g [U,U^\prime]}$ is in $p p \to \gamma\, {\rm jet}\, X$, if one can select a kinematic region where the partonic subprocess $qg \to \gamma q$ dominates over $q\bar{q} \to \gamma g$. In this case one can access $f_1^{g[+,-]}$. Such a study has already been proposed in $p A$ scattering in~\cite{Dominguez:2011wm,Schafer:2013mza}. Large $A$ and small $x$ (high energy) help to select the gluon induced subprocesses and lead to simplifications regarding the study of the process dependence of the gluon TMD $f_1^{g [U,U^\prime]}$~\cite{Xiao:2010sp,Xiao:2010sa,Dominguez:2011wm}. In several of the mentioned processes also 
$h_1^{\perp g [U,U^\prime]}$ can be accessed \cite{Boer:2011kf,Boer:2010zf,Boer:2014lka} with the same link structure as 
$f_1^{g [U,U^\prime]}$. A study of $h_1^{\perp g [U,U^\prime]}$ in $eA$ and $pA$ collisions has been performed in~\cite{Metz:2011wb}.
 
The average transverse momentum $\langle p_\st^2 \rangle$ and its broadening $\Delta p_\st^2 \equiv \langle p_\st^2 \rangle_A - \langle p_\st^2 \rangle_p$ with atomic number $A$ has been studied extensively in the literature in terms of the collinear factorization approach at twist-4~\cite{Luo:1992fz,Guo:1998rd,Kang:2008us,Kang:2011bp,Xing:2012ii,Kang:2013ufa}. The relevant twist-4 parton distribution functions~\cite{Jaffe:1983hp,Qiu:1988dn} are (at tree level) related to the first transverse moment $f_1^{(1)}(x)$ for quarks and gluons, but also determine the large-$p_\st$ perturbative tail of the TMDs $f_1^{[U]}$ and $f_1^{g [U,U^\prime]}$. The collinear twist-4 functions involve only light-cone operators, all having the GLs with finite paths along the light-cone and are thus process-independent. In the twist-4 calculations the process dependence is attributed to the hard partonic scattering factors. This implies process-dependent relations between the TMDs $f_1^{[U]}$ and $f_1^{g [U,U^\prime]}$ and the twist-4 functions, in analogy to the relation between the Sivers and Qiu-Sterman functions~\cite{Boer:2003cm}. See Ref.~\cite{Xing:2012ii} for a detailed discussion of the process dependence of nuclear broadening in the twist-4 approach.

To conclude, in this paper we have elucidated the operator structure of quark and gluon correlators relevant to situations where transverse momenta show up at the quadratic level, including as examples the quadratic $p_\st$ dependence but including in this both the $p_\st^2$ dependence of the functions as well as the rank 2 tensor combination multiplying these functions. From this analysis it becomes clear which parts of the $p_\st$ dependence of TMDs give process dependence. This is achieved by splitting the TMDs into universal operator combinations and process-dependent coefficients. This is of relevance (among other cases) for the $p_\st$-width of TMDs which enters in $p_\st$-broadening observables, for observables involving pretzelocity functions for transversely polarized quarks in a transversely polarized nucleon, and for observables sensitive to linearly polarized gluons in an unpolarized nucleon. For the unpolarized gluon TMD $f_1^{g [U,U^\prime]}$ we have discussed ways to experimentally test the nonuniversality, in both electron-proton and proton-proton collisions. It would be very interesting if it could be established experimentally that even such a T-even TMD shows process dependence due to the gauge link structure. As it may be challenging to achieve this goal, we propose a way to study the gauge link dependence of $f_1^{[U]}$ quantitatively using lattice QCD computations.

\section*{Acknowledgements}
We acknowledge discussions with Markus Diehl. This research is part of the research program of the ``Stichting voor Fundamenteel Onderzoek der Materie (FOM)", which is financially supported by the ``Nederlandse Organisatie voor Wetenschappelijk Onderzoek (NWO)" as well as the FP7 "Ideas" programme QWORK (Contract No. 320389).


\end{document}